# Snooker Statistics and Zipf's Law

## Wim Hordijk

*Email*: wim@WorldWideWanderings.net
*ORCiD:* 0000-0002-0223-6194

**Abstract**
Zipf's law is well known in linguistics: the frequency of a word is inversely proportional to its rank. This is a special case of a more general *power law*, a common phenomenon in many kinds of real-world statistical data. Here, it is shown that snooker statistics also follow such a mathematical pattern, but with varying (estimated) parameter values. Two types of rankings (prize money earned and centuries scored), and three time-frames (all-time, decade, and year) are considered. The results indicate that the power law parameter values depend on the type of ranking used as well as the time-frame considered. Furthermore, in some cases the resulting parameter values vary significantly over time, for which a plausible explanation is provided.

## Introduction

Zipf's law is well known in linguistics (Zipf 1935, 1949; Piantadosi 2014). When words are ranked according to their frequency of use (from highest to lowest), the frequency *f(r)* of a word turns out to be inversely proportional to its rank *r*. In other words, the most frequent word (*r*=1) occurs roughly twice as often as the second-most frequent word (*r*=2), three times as often as the third-most frequent word (*r*=3), and so on.

This inverse relationship is a special case of a more general mathematical *power law*

$$f(r) \propto \frac{1}{r^{\alpha}}$$

with the parameter *α* being (close to) one in the case of Zipf's law.

However, such power laws are not restricted to languages. They show up in many different situations, such as academic citations, website hits, earthquake magnitudes, intensity of solar flares, city sizes, wealth distributions (Newman 2005), molecular networks (Cazzolla Gatti et al. 2018), technological innovation (Steel et al. 2020), and many more. Indeed, they seem to be a universal phenomenon (Corominas Murtra and Solé 2010). It is not surprising, then, that they also show up in sports rankings (Deng et al. 2012; Morales et al. 2016, 2021).

Deng et al. (2012) showed that power laws can be found in ranking statistics across a range of different sports, including snooker. However, they used cumulative distributions rather than explicit rank distributions. Furthermore, they only considered statistics from one particular point in time.

Morales et al. (2016, 2021) confirm power law occurrences in ranking statistics of various sports (although not including snooker), using actual rank distributions. They also study a dynamic aspect of rankings, in terms of what they call rank diversity (a measure of the number of players occupying a given rank during a particular time period). However, they do not investigate if or how the parameter *α* of the power laws change over time.

Building on an earlier preliminary investigation (Hordijk 2019), a more detailed and comprehensive study of the occurrence of power laws in snooker statistics is presented here. First, power laws are shown to occur in two different player ranking statistics: total prize money earned, and number of centuries scored. Next, it is shown that there is a difference between the estimated power law parameter values for these two ranking types, and also between all-time statistics and shorter time-frame statistics (one year or one decade). Finally, it is investigated whether these estimated parameter values change over time, and what might cause such variations.

**Methods**

A power law shows up as a straight line in a log-log plot. In such a plot, both axes are on a logarithmic scale, rather than a linear scale. So, as a first indication of whether a particular data set might follow a power law, this data can be visualized in a log-log plot to see how closely the data points fall along a straight line.

An example is given in Figure 1, using the 50 most frequent words from the *Corpus of Contemporary American English*, or COCA (Davies 2022). The COCA contains more than one billion words from contemporary English texts, spanning many different literature categories and authors. This data appears to fall along a straight line quite accurately.

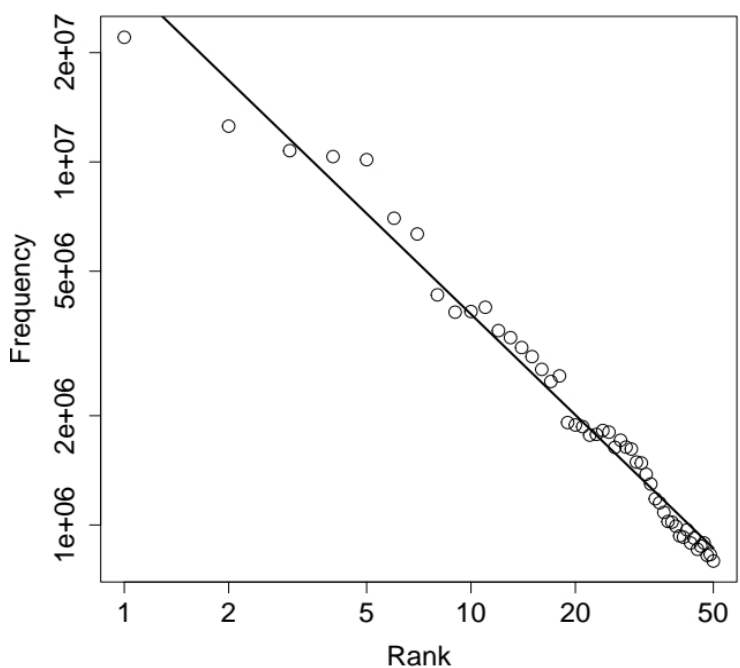

*Figure 1: A log-log plot of word frequency against rank for the 50 most frequent words from the COCA. The solid line represents an estimated power law fit.*

Next, a power law can be estimated to fit the data by performing a linear regression on the logarithms of the rank and frequency values. This fit is represented by the solid line in the plot, and results in an estimated parameter value $α = 0.922$. This value is indeed close to one, as originally noted by Zipf more than 80 years ago. The value of $α$ actually determines the slope of the straight line representing the power law in a log-log plot. A larger value of $α$ results in a steeper line, a smaller value in a shallower line.

A linear regression also provides a "goodness-of-fit" measure ($R^2$), a value between zero and one. The closer to one, the better the fit. As expected, the fit in the example above is very good: $R^2 = 0.98$.

Real-world data never falls *exactly* along a straight line though, partly because of finite size effects. An ideal power law assumes that arbitrarily large frequencies are possible, and that the sample size is large enough so that rare events are always observed. Neither of these is hardly ever the case with real data, so there will always be deviations from a perfect straight line, especially at the top of the ranking (upper left in the plot).

Adjustments to the basic power law formula can be made (such as adding an exponential decay term) to take these finite size effects into account and improve the fit somewhat. However, the primary goal here is not so much to account for every little deviation from a straight line, but to compare the (estimated) values of the main power law parameter $\alpha$ between different types of rankings and across time. Therefore, the basic power law formula (as shown in the introduction above) is used here, which is fit to the data with a linear regression on the logarithmic values. All statistical analyses were done using the R language for statistical computing (R Core Team 2021).

The snooker statistics used here consist of two types of professional player rankings: total prize money earned, and number of centuries scored (a century in snooker is a score of 100 or more points in a single visit to the table). Three different time-frames are considered: all-time, decade, and year. The all-time rankings were current as of January 2022, and the past decade (2010-2019) and past year (2021) were considered for the shorter time-frames. Furthermore, to investigate how the resulting power laws may have changed over time, data for the two previous decades (1990-1999 and 2000-2009), and for the years 1990, 1995, 2000, 2005, 2010, 2015, and 2020 were also included. All ranking data was obtained from `CueTracker` (Florax 2022). For a fair comparison across the different ranking types and time-frames, the data sets were ensured to all be of the same size by taking the top 100 entries of each ranking.

## Results

Figure 2 shows the snooker statistics in a log-log plot for the two rankings based on prize money (left) and centuries (right), and for the three time-frames of all-time (open circles), past decade (closed circles), and past year (open squares). The solid lines represent estimated power laws to fit the data.

The resulting estimated parameter values $\alpha$ and goodness-of-fit values $R^2$ are presented in Table 1, for the two rankings and three time-frames.

|          | Prize money | | Centuries | |
|----------|-------------|-------|-----------|-------|
|          | $\alpha$    | $R^2$ | $\alpha$  | $R^2$ |
| All-time | 0.857       | 0.96  | 0.741     | 0.94  |
| Decade   | 0.984       | 0.95  | 0.837     | 0.92  |
| Year     | 0.978       | 0.96  | 0.863     | 0.94  |

*Table 1: Estimated parameter and goodness-of-fit values for the power laws fitted to the snooker statistics.*

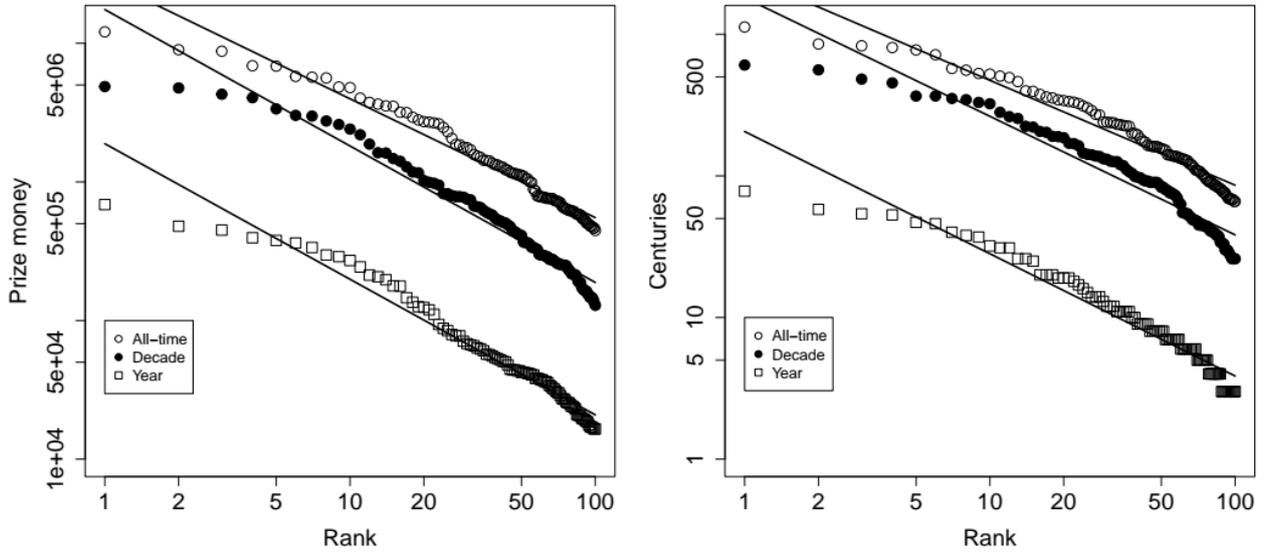

*Figure 2: The snooker data for the two ranking types and three time-frames. Solid lines represent estimated power laws.*

Figure 3 (left) shows how the estimated parameter values *α* change over time, for the two rankings based on prize money (solid lines) and centuries (dashed lines), and the two time-frames of a year (open squares) and a decade (closed circles).

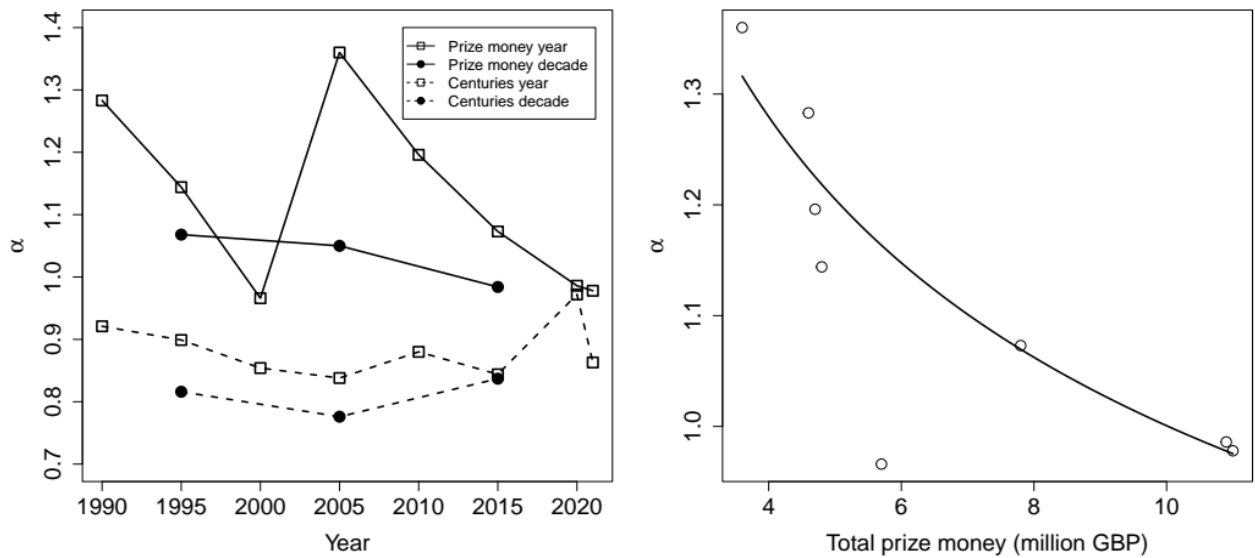

*Figure 3: Left: The change in estimated power law parameter over time. Right: The estimated parameter values for the prize money ranking in a given year, against the total amount of prize money available that year.*

Figure 3 (right) shows the values of α for the prize money ranking in a given year, against the total amount of prize money available, combined over all tournaments that year, in million GBP (this plot is explained in more detail below in the discussion section).

## Discussion

As Figure 2 clearly shows, power laws do indeed show up in snooker statistics. However, the finite size effects (i.e., deviations from a straight line in a log-log plot) are more pronounced than in the word frequency example of Figure 1. This is also reflected in the $R^2$ values (Table 1), which range from 0.92 to 0.96 for snooker statistics, compared to 0.98 for word frequencies.

However, this slightly lower goodness-of-fit is still quite acceptable, and not surprising given that the word frequencies are based on a corpus (the COCA) of more than one billion words. In contrast, the number of snooker tournaments played is around 30 per year, or 300 per decade. So, even with the all-time rankings based on several thousand tournaments, that is still only a tiny fraction of the size of the COCA.

What Figure 2 and Table 1 also show, is that there is a clear distinction between the estimated power law parameter values α of the all-time rankings on the one hand, and the rankings over shorter time-frames such as a year or decade on the other hand. For both ranking types (prize money and centuries), the all-time rankings result in a smaller value of α than for the shorter time-frames.

Similarly, there is a clear distinction between the two ranking types themselves, with the estimated values of α being consistently larger (given the time-frame considered) for the prize money ranking than for the centuries ranking. Also note that for the prize money ranking and the decade and year time-frames, the estimated α values are very close to one, as in the original Zipf's law.

Finally, as Figure 3 (left) shows, the estimated parameter values might fluctuate over time. Although in most cases they appear to be fairly robust, the estimated values of α vary quite significantly for the prize money rankings on a yearly basis.

To find a possible explanation for this behavior, the total amount of prize money available, combined over all tournaments in a given year, was also obtained from `CueTracker` (Florax 2022) for the years considered here. The estimated value of α for a given year was then plotted against the total amount of prize money (in million GBP) available in that year. This plot is shown in Figure 3 (right).

Clearly, a smaller amount of total prize money results in a larger value of α, and vice versa. There is one outlier, near the bottom-left of the plot (with a total prize money of close to 6 million GBP, and a value of α below 1.0). Interestingly, though, when ignoring this outlier, the remaining points seem to follow a power law as well.

To check this, a power law was estimated for this data (minus the outlier), the result of which is represented by the solid curve in Figure 3 (right). Note that this plot is on a regular (linear) scale, so the power law does not show up as a straight line, but is curved. The estimated power law seems to be a fairly reasonable fit ($R^2 = 0.92$), and thus provides a plausible explanation for the variation in the estimated values of α for the prize money ranking on a yearly basis. The total amount of available prize money obviously constrains the possible distribution of prize money among the players.

Variations in the parameter *α* are interesting in the sense that this parameter indicates how much a ranking is dominated by the top players. This relates to the well-known "80-20 rule", which says that, for example, 20% of the people own 80% of the wealth in a society. However, with a power law it depends on the value of *α* what the actual percentage of the top players is that is responsible for 80% of the prize money earned, or centuries scored.

When *α* = 1, as in Zipf's law, it is actually the top 35% of the ranking that earns 80% of the prize money, or scored 80% of all centuries (assuming there are 100 players in the ranking). However, when *α* > 1 (corresponding to a steeper line in the log-log plot), the ranking is more dominated by the top players. For example, for *α* = 1.3, just the top 15% of the players together already earn 80% of the prize money. In contrast, when *α* < 1 (corresponding to a shallower line in the log-log plot), the distribution becomes somewhat more equal. For example, for *α* = 0.7, it takes more than half (56%) of the top players to earn 80% of the prize money.

## Conclusions

In conclusion, although power laws indeed appear to be common in snooker statistics, their particular form (i.e., parameter value, or slope) depends strongly on the type of ranking used as well as the time-frame considered. Furthermore, this parameter value may change over time, and a plausible explanation for such behavior was provided in terms of the total amount of resources (in this case prize money) available. Finally, the value of the parameter *α* indicates how much a ranking is dominated by the top players. The larger the value of *α*, the more the top players dominate the rankings and total earnings.

It would be interesting to see if the results obtained here can be observed in the ranking statistics of other sports as well. Deng et al. (2012) and Morales et al. (2016, 2021) do provide estimated power law parameter values for a range of different sports, but did not analyze to what extent these depend on ranking type used or time-frame considered, or whether these values change over time. Axtell (2001) presents estimated power law parameter values for US firm size statistics measured over the span of one decade. However, those values seem very robust and do not change at all over time. Further investigation into this for general sports statistics would be useful.

Although power laws are a common phenomenon that can be generated by many different types of processes (Newman 2005), Deng et al. (2012) and Morales et al. (2016, 2021) suggest possible mechanisms specific to competitive sports. Using computer simulations, they then confirmed the emergence of power laws from these specific mechanisms, which might therefore also be relevant to snooker statistics. Furthermore, O'Brien and Gleeson (2021) propose an alternative way to analyze snooker player rankings, based on complex networks. These studies all suggest useful directions for further research.

As a final note, in the year 1935, when Zipf first described his findings of an inverse relationship between word frequency and rank (Zipf 1935), there was only one professional snooker tournament held: the world championship. Just five players competed in the tournament, and there was no prize money (apparently the players made some money from spectator ticket sales). Even if Zipf had been interested in snooker, he certainly would not have found his famous relationship in such scant statistics. But now, almost 90 years later, power laws clearly abound in snooker rankings.

## Acknowledgments
The research and results presented in this article were motivated by watching this year's (2022) Masters snooker tournament. No funding was received to do this work, and the author declares no conflict of interest in any way.

## References

Axtell, R. L. (2001). Zipf distribution of U.S firm sizes. *Science* **293**:1818–1820.

Cazzolla Gatti, R., Fath, B., Hordijk, W., Kauffman, S. A., and Ulanowicz, R. (2018). Niche emergence as an autocatalytic process in the evolution of ecosystems. *Journal of Theoretical Biology* **454**:110–117.

Corominas Murtra, B. and Solé, R. V. (2010). Universality of Zipf's law. *Physical Review E* **82**:011102.

Davies, M. (2022). *The Corpus of Contemporary American English (COCA)*. www.english-corpora.org/coca

Deng, W., Li, W., Cai, X., Bulou, A., and Wang, Q. A. (2012). Universal scaling in sports ranking. *New Journal of Physics* **14**:093038.

Florax, R. (2022). *CueTracker.* CueTracker.net

Hordijk, W. (2019). The power of snooker. *Plus magazine*, April 2019. plus.maths.org/content/power-snooker

Morales, J. A., Flores, J., Gershenson, C., and Pineda, C. (2021). Statistical properties of rankings in sports and games. *Advances in Complex Systems* **24**:2150007.

Morales, J. A., Sánchez, S., Flores, J., Pineda, C., Gershenson, C., Cocho, G., Zizumbo, J., Rodriguez, R. F., and Iñiguez, G. (2016). Generic temporal features of performance rankings in sports and games. *EPJ Data Science* **5**:33.

Newman, M. E. J. (2005). Power laws, Pareto distributions and Zipf's law. *Contemporary Physics* **46**:323–351.

O'Brien, J. D. and Gleeson, J. P. (2021). A complex networks approach to ranking professional snooker players. *Journal of Complex Networks* **8**:cnab003.

Piantadosi, S. T. (2014). Zipf's word frequency law in natural language: A critical review and future directions. *Psychonomic Bulletin & Review* **21**:1112–1130.

R Core Team (2021). *R: A language and environment for statistical computing*. www.R-project.org

Steel, M., Hordijk, W., and Kauffman, S. A. (2020). Dynamics of a birth-death process based on combinatorial innovation. *Journal of Theoretical Biology* **491**:110187.

Zipf, G. K. (1935). *The Psychobiology of Language*. Houghton-Mifflin, New York, NY.

Zipf, G. K. (1949). *Human Behavior and the Principle of Least Effort*. Addison-Wesley, Cambridge, MA.